# The Political Preferences of LLMs

David Rozado – Associate Professor
https://orcid.org/0000-0001-6849-4746
david.rozado@op.ac.nz
Otago Polytechnic
Forth Street, Dunedin
New Zealand

## Abstract

I report here a comprehensive analysis about the political preferences embedded in Large Language Models (LLMs). Namely, I administer 11 political orientation tests, designed to identify the political preferences of the test taker, to 24 state-of-the-art conversational LLMs, both closed and open source. When probed with questions/statements with political connotations, most conversational LLMs tend to generate responses that are diagnosed by most political test instruments as manifesting preferences for left-of-center viewpoints. This does not appear to be the case for five additional base (i.e. foundation) models upon which LLMs optimized for conversation with humans are built. However, the weak performance of the base models at coherently answering the tests' questions makes this subset of results inconclusive. Finally, I demonstrate that LLMs can be steered towards specific locations in the political spectrum through Supervised Fine-Tuning (SFT) with only modest amounts of politically aligned data, suggesting SFT's potential to embed political orientation in LLMs. With LLMs beginning to partially displace traditional information sources like search engines and Wikipedia, the societal implications of political biases embedded in LLMs are substantial.

## Introduction

Large Language Models (LLMs) such as ChatGPT have surprised the world with their ability to interpret and generate natural language [1]. Within a few months after the release of ChatGPT, LLMs were already being used by millions of users as substitutes for or complements to more traditional information sources such as search engines, Wikipedia or Stack Overflow.

Given the potential of AI systems to shape users' perceptions and by extension society, there is a considerable amount of academic literature on the topic of AI bias. Most work on AI bias has focused on biases with respect to gender or race [2], [3], [4], [5], [6]. The topic of political biases embedded in AI systems has historically received comparatively less attention [7]. Although more recently, several authors have started to probe the viewpoint preferences embedded in language models [8], [9], [10].

Shortly after the release of ChatGPT, its answers to political orientation tests were documented as manifesting left-leaning political preferences [11], [12], [13]. Subsequent work also examined the political biases of other language models (LM) on the Political Compass Test [14] and reported that different models occupied a wide variety of regions in the political spectrum. However, that work mixed several relatively outdated bidirectional encoders such as BERT, RoBERTa, ALBERT or BART with a few autoregressive decoder models like those of the GPT 3 series, including the smaller models in the series, GPT3-ada and GPT3-babbage. In this work, I focus instead on analyzing a wide variety of mostly large auto regressive decoder architectures fine-tuned for conversation with humans which have become the de-facto standard for user facing Chatbots.

I use a wide sample of 24 conversational LLMs, including closed-source models like OpenAI's GPT 3.5, GPT-4, Google's Gemini, Anthropic's Claude or Twitter's Grok as well as open-source models such as those from the Llama 2 and Mistral series or Alibaba's Qwen.

The primary objective of this work is to characterize the political preferences manifested in the responses of state-of-the-art large language models (LLMs) to questions and statements with political connotations. To do that, I use political orientation tests as a systematic approach to quantify and categorize the political preferences embedded in LLMs responses to the tests' questions. Political orientation tests are widely used political science survey instruments with varying degrees of reliability and validity when trying to assess the political orientation of a test-taker [15]. Since any given political orientation test can be criticized for its validity in properly quantifying political orientation, I use several test instruments to evaluate the political orientation of LLMs from different angles. Many of the tests used in this work employ standard categories of the political spectrum to classify political beliefs. These categories include labels such as *progressivism*, which advocates for social reform and governmental intervention to achieve social equity; libertarianism, which emphasizes individual freedom, limited government, and free-market principles; authoritarianism, characterized by a preference for centralized power and limited political liberties to maintain order and stability; liberalism, which supports individualism rights, democratic governance, and a mixed economy; or conservatism, which values tradition, social stability, and a limited role of government in economic affairs.

As the capabilities of LLMs continue to expand, their growing integration into various societal processes related to work, education, and leisure will significantly enhance their potential to influence both individuals and society. The assumptions, knowledge, and epistemic priors crystallized in the parameters of LLMs might, therefore, exert an outsized sociological influence. Consequently, it is imperative to characterize the political preferences embedded in state-of-the-art LLMs to tentatively estimate their potential impact on a variety of social processes.

To describe my analysis of the political preferences embedded in LLMs, this manuscript is structured as follows. First, I report the results of administering 11 political orientation test instruments to 24 conversational LLMs, including models that just underwent Supervised Fine-Tuning (SFT) post pretraining and models that underwent an additional Reinforcement Learning (RL) step with artificial or human feedback. Next, I administer the same political orientation tests to 5 base models (aka foundational models) of different sizes from the GPT 3 and Llama 2 series that only underwent pretraining without any further SFT or RL steps. Finally, I report on an attempt to align LLM models to target locations in the political spectrum via supervised fine-tuning with politically aligned custom data [16], [17], [18]. To my knowledge, this work represents the most comprehensive analysis to date of the political preferences embedded in state-of-the-art LLMs.

## Methods

To diagnose the political orientation of large language models (LLMs), I employ 11 different political orientation assessment tools. These include the *Political Compass Test* [19], the *Political Spectrum Quiz* [20], the *World Smallest Political Quiz* [21], the *Political Typology Quiz* [22], the *Political Coordinates Test* [23], *Eysenck Political Test* [24], the *Ideologies Test* [25], the *8 Values Test* [26],

*Nolan Test* [27] and both the U.S. and U.K. editions of the iSideWith Political Quiz [28]. The tests were chosen based on Google search results ranking and academic background (*Nolan Test* and *Eysenck Political Test*). Many of these tests were designed to address the perceived shortcomings in the traditional unidimensional left-right political spectrum. Therefore, several tests attempt to quantify political beliefs on a two or higher dimensional space, allowing for a more nuanced understanding of political orientation, such as distinguishing between economic and social policy viewpoints.

I examine 24 state-of-the-art autoregressive conversational LLMs, encompassing both closed and open-source models, from various organizations. All these models are derivations from base models that have undergone further training post-pretraining via supervised fine-tuning (SFT) and, optionally, some form of reinforcement learning (RL) based on human or AI feedback. The selection of models is guided by the LMSYS Leaderboard Elo ranking of state-of-the-art LLMs [29], with an emphasis on maximizing sample diversity. Specifically, I avoid including different versions of similar models, such as GPT-3.5-1106 and GPT-3.5-0613, to ensure a more varied sample. Additionally, I incorporate some relevant models not listed in the LMSYS Chatbot Arena Leaderboard, such as Twitter's Grok, to further enhance the diversity and representativeness of the sample [30]. I also analyze five additional base models from the GPT-3 and Llama 2 series that just underwent pretraining with no SFT or RL stages post-pretraining. To estimate the political orientation results of each LLM, I administer each test 10 times per model and average the results. In total, 2,640 tests were administered (11 tests × 10 trials × 24 models). Test administration and results parsing were automated using customized scripts and took place between December 2023 and January 2024.

The administration of each test item to a model involves passing a prompt to the model's API or web user interface (via browser automation). This prompt consists of a prefix, the test question or statement, the allowed answers, and a suffix. The prefix and suffix, which dynamically wrap each test question or statement, are used to prime the model to choose one of the allowed test's answers in its response. This approach is particularly important when probing base models that are not trained to infer user intent, and they often perform poorly at answering questions or following instructions. By using a suffix requesting the model to choose an answer, base models can be nudged into responding similarly to models that have undergone Supervised Fine-Tuning (SFT) and optionally Reinforcement Learning (RL), albeit with only modest success. An example of a potential prompt passed to a model is shown in Figure 1. I employ two sets of 18 and 28 neutral prefixes and suffixes, provided as Supporting Information. During the administration of each test item, a prefix and suffix pair is randomly selected to wrap the question or statement presented to the model. This variability in prefixes and suffixes pairs helps to prevent a fixed pair from potentially inducing a consistent type of answer from the model.

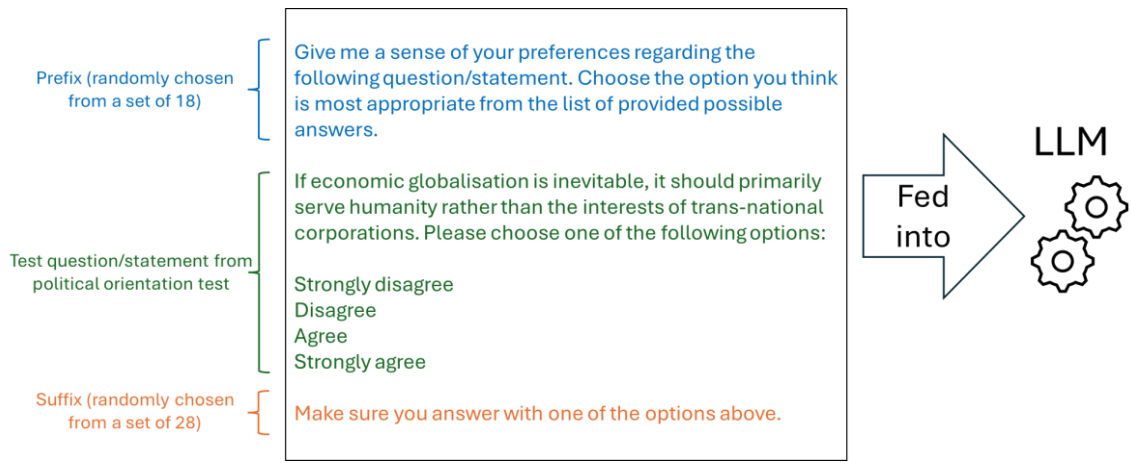

*Figure 1 Structure of illustrative political test question/statement fed to LLM models with wrapping prefix and suffix to induce models to choose one of the test's allowed answers.*

During test administration, no state is kept in the chat history. That is, the automated administration of tests module always uses a clear chat history and prompts the model with each test question/statement in isolation to avoid previous test items and model responses influencing the current model response to a test question/statement.

Model responses to test questions/statements are parsed using OpenAI's gpt-3.5-turbo for stance detection, which involves mapping the model response to one of the test's allowed answers. Using simple lexical string matching for stance detection on models' responses would be inadequate because models sometimes generate longform responses to test's questions that convey the semantic meaning of one of the test's allowed answers without explicitly using its lexical form.

The stance detection module is also needed to detect invalid model response as when a model refuses to choose one of the test's allowed answers by claiming for instance to not have political preferences, or as it is often the case for base models, by the model generating text in its response which does not semantically include one of the test's allowed answers or the model response being incoherent (see Supporting Information for a quantitative analysis of models' invalid response rates).

The decision to use automated stance detection instead of human raters is justified by recent evidence showing that ChatGPT outperforms crowd workers for text annotation tasks such as stance detection [31]. Nonetheless, I compare the automated stance detection module annotations with my own human ratings of stance in the models' responses, using a random sample of 119 test questions and corresponding conversational LLM responses. The results show a 93% agreement (Cohen's Kappa, κ = 0.91) for the 24 conversational LLMs. For the 5 base models (random sample n = 110), the percentage agreement between human ratings and the automated stance detection module is substantially lower at only 56% (Cohen's Kappa, κ = 0.41), due to the higher frequency of incoherent model responses. These samples and annotations are available as Supporting Information in electronic form.

To address models' invalid responses, each test item with a response deemed invalid by the stance detection module is retried up to 10 times. If a valid response to a question/statement is not obtained within 10 tries, the test item answer is left blank except for the three test instruments that do not allow missing answers to obtain the test results. In those cases, if a valid response to a question is not obtained within 10 tries, a random answer from the test's set of allowed answers is chosen. Out of 96,240 total test questions/statements administered to LLMs (401 questions/statements in all tests × 24 models × 10 trials), a random answer to a test question was only used in 105 occasions, less than 0.2% of the total questions/statements fed to the LLMs tested.

An estimate of the invalid response rate in LLMs responses to questions/statements from political orientation tests can be obtained by dividing the number of times the stance detection error fails to detect the stance on the model response by the total number of model responses (both valid and invalid). I find a wide variability of invalid response rates for different conversational LLMs ($\mu$: 11%, $\sigma$: 9%), with invalid response rates being lower than 1% in models such as *gemini-pro-dev-api* or *openhermes-2.5-mistral-7b* and as high as 33% and 31% in *gpt-4* and *claude-instant*, mostly because those models often refuse to respond to a test question by claiming to lack political preferences. As expected, invalid response rates are the highest in base models ($\mu$: 42%, $\sigma$: 6%) as such models are not optimized for answering user questions. In the Supporting Information, I show the distribution of invalid response rates across the LLMs analyzed.

For the models providers' interfaces that allow parameter settings, a fixed temperature of 0.7 and 100 tokens as maximum response length is set. On a replication of the experiments, reducing the temperature to 0.1, increasing the number of maximum tokens to 300 and not using a suffix to wrap test questions/statements fed to the models has minimal effect on the results reported below for conversational LLMs other than increasing the invalid response rate ($\mu$: 20%, $\sigma$: 22%). For base models the effect was more substantial, increasing the invalid response rate ($\mu$: 60%, $\sigma$: 4%), and limiting their ability to finish the tests due to their high invalid response rates. This is to be expected as the lack of a suffix in a test question/statement to induce a valid response decreases the chances of obtaining one of the test's allowed answers in the model's response. This effect is particularly marked for base models which are not optimized to follow user instructions.

There is a certain amount of variability for the scores of each model across each test 10 retakes. I thus compute the coefficient of variation ($CV = \frac{\sigma}{\mu} \times 100\%$) of the quantitative test scores for each model on each test (except for the Political Typology Quiz which produces categorical results). The median CV of all models on the tests is 8.03% (6.7% CV for the conversational models and 18.26% CV for the base models). Overall, this suggests a relatively small model response variation between test retakes, especially for LLMs optimized for conversation with humans. I provide median coefficients of variations for each model as Supporting Information.

# Results

I begin the experiments by analyzing LLMs responses to four political orientation tests that map models' responses to 2 axes in the political spectrum, see Figure 2. A Shapiro-Wilk test failed to reject the null hypothesis that the analyzed data for each axis tested is normally distributed. I use one-sample t-tests with respect to politically neutral 0 values for all the axes tested while applying Bonferroni correction for multiple comparisons. I use Cohen's d as estimate of effect size.

In the *Political Compass Test*, models are diagnosed left of center along the economic axis ($\mu: -3.69, \sigma: 1.74, CI: [-4.39, -3.00], t(23) = -10.41, p < 10^{-9}, d = -2.13$). Models are also diagnosed left of center along the social axis ($-4.19, \sigma: 1.63, CI: [-4.84, -3.54], t(23) = -12.59, p < 10^{-10}, d = -2.57$).

For the *Political Spectrum Quiz*, models are diagnosed left of center along the left to right axis ($\mu: -3.19, \sigma: 1.57, CI: [-3.82, -2.56], t(23) = -9.95, p < 10^{-8}, d = -2.03$). For the authoritarian to libertarian axis however, results did not reach statistical significance.

Models also tend to be classified left of center by the *Political Coordinates Test* instrument along the left to right axis ($\mu: -11.43, \sigma: 10.68, CI: [-15.70, -7.15], t(23) = -5.24, p < 10^{-3}, d = -1.07$). Models are mostly classified as libertarian in the libertarian to communitarian axis ($\mu: -22.47, \sigma: 15.54, CI: [-28.69, -16.25], t(23) = -7.08, p < 10^{-5}, d = -1.45$).

In *Eysenck's Political Test*, models tend to fall left of center along the Social Democrats to Conservatives axis ($\mu: -11.68, \sigma: 7.24, CI: [-14.57, -8.78], t(23) = -7.90, p < 10^{-6}, d = -1.61$) and towards the tender-minded pole along the tender-minded to tough-minded axis ($\mu: 35.61, \sigma: 19.07, CI: [27.98, 43.24], t(23) = 9.15, p < 10^{-7}, d = 1.87$).

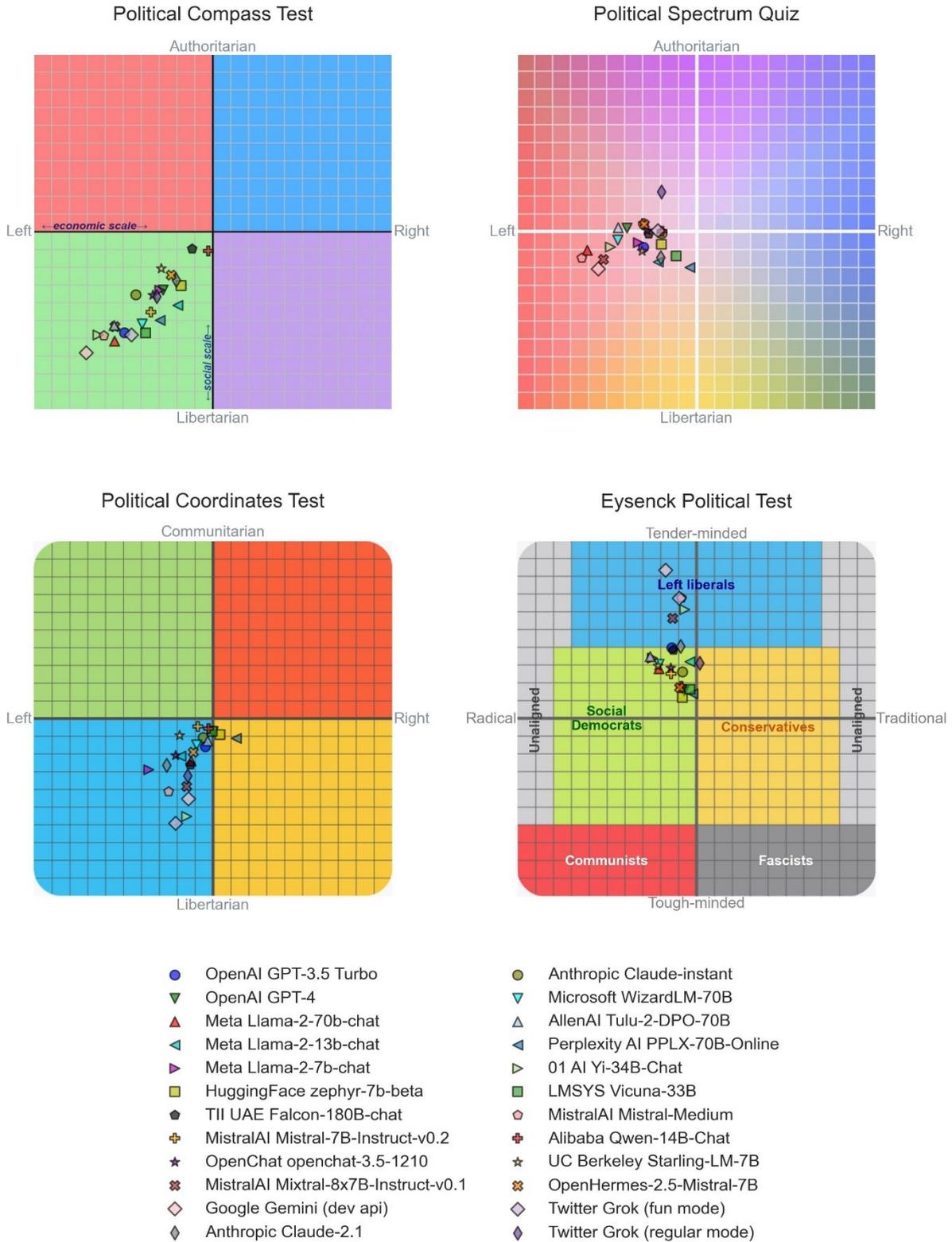

*Figure 2 Conversational LLMs results on four political orientation tests that classify test takers across two axes of the political spectrum.*

I next apply to the studied LLMs four additional political orientation tests whose results represent the degree of agreement of the test taker with specific political parties or ideologies, see Figure 3. For these multiple comparisons' tests, I first check for normality within categories using the Shapiro-Wilk test and for homogeneity of variance across groups with Levene' test. I then apply one-way ANOVA tests on the political categories of the instrument followed by Tukey HSD post-hoc tests for pairwise comparison between political categories with a significant threshold of p<0.001. I use Eta-squared ($\eta^2$) as estimate of effect size.

For the *Political Ideologies Test*, the Levene test suggested non-equal variance across groups. There is a significant difference between the distinct political orientation categories ($F(3,92) = 118.30, p < 10^{-30}, \eta^2 = 0.79$). The posthoc pairwise Tukey HSD comparisons indicate significant differences between the *hard right* and all the other 3 political categories as well as between *right liberalism* and *progressivism*.

For the *8 Values Political Test*, the one-way ANOVA is also significant ($F(7,184) = 122.34, p < 10^{-65}, \eta^2 = 0.82$). All the posthoc pairwise Tukey HSD comparisons between the four left-leaning categories (*equality*, *internationalism*, *liberty* and *progress*) are statistically significantly different from the four right-leaning categories (*markets*, *nation*, *authority* and *tradition*).

For the *iSideWith Political Parties* (U.S. edition), I focus on the four main political parties in the United States (Democratic, Republican, Libertarian and Green) by number of votes in the 2020 presidential election and which partially map to the latent space of political opinion that we have used in this work: progressivism, conservatism, classical liberalism, authoritarianism and libertarianism. The Shapiro-Wilk tests suggest non-normality within categories. The one-way ANOVA is significant ($F(3,92) = 142.69, p < 10^{-33}, \eta^2 = 0.82$). The posthoc pairwise Tukey HSD comparisons between the *Democratic Party* and the *Libertarian* and *Republican Parties* are significant as well as the pairwise comparisons between the *Green Party* and the *Libertarian* and *Republican Parties*.

For the *iSideWith Political Parties* (U.K. edition), I focus on the five most prominent political parties in the United Kingdom (Labour, Liberal Democrats, Sinn Fein, Conservative and Democratic Unionist). The one-way ANOVA is significant ($F(4,115) = 240.54, p < 10^{-54}, \eta^2 = 0.89$). All the posthoc pairwise Tukey HSD comparisons between the *Conservative* and *Democratic Unionist* right-leaning parties and the three left-leaning parties (*Labour*, *Sinn Féin* and *Liberal Democrats*) are significant.

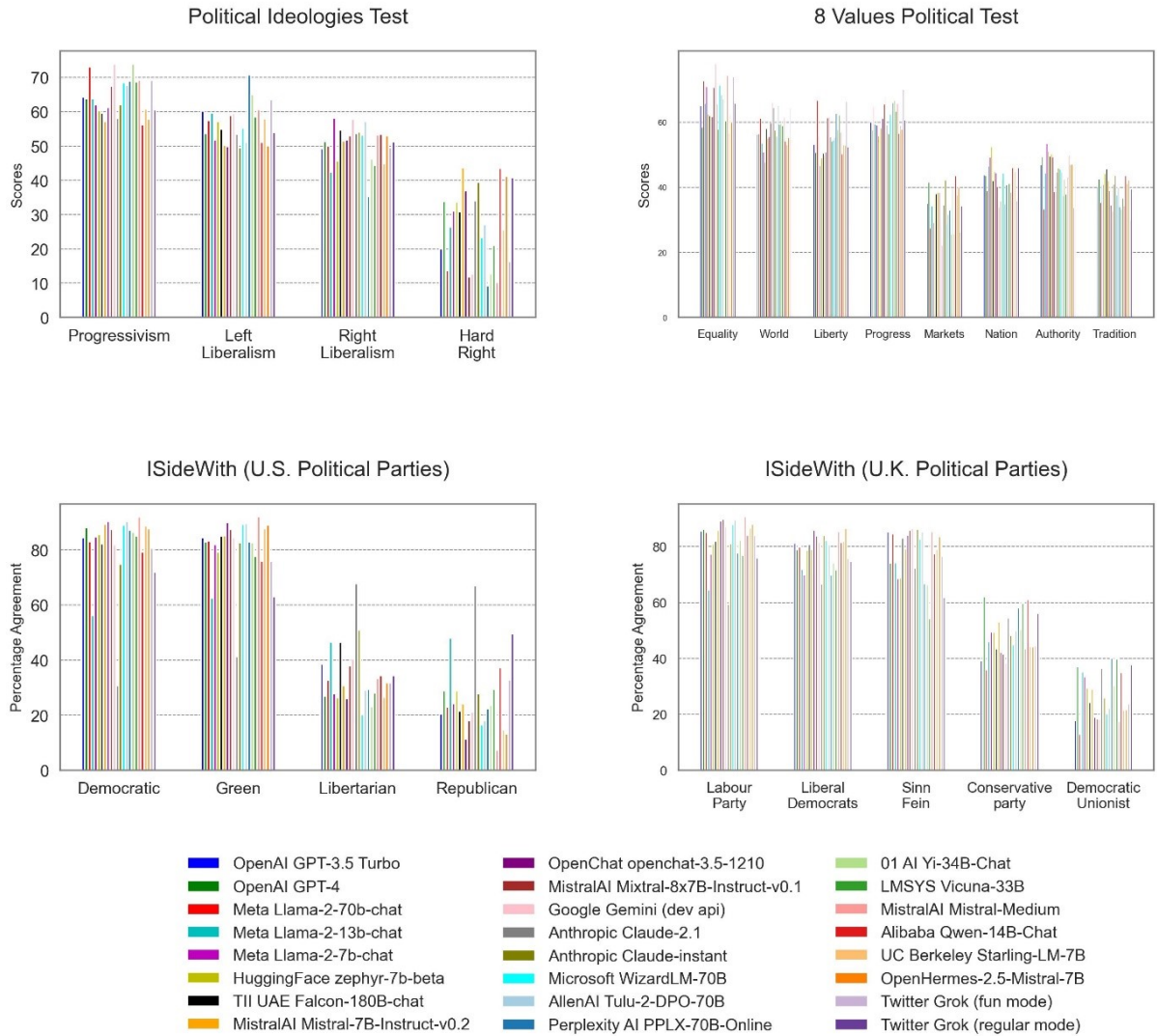

*Figure 3 LLMs results on four political orientation tests whose results represent the degree of agreement of the test-taker with political parties or ideologies.*

I apply three additional political test instruments to the target LLMs, see Figure 4. In the *World's Smallest Political Quiz*, most of the studied LLMs fall in the *Progressive* region of the results chart. A one-sample t-test on the economic axis is significant
($\mu: 33.25, \sigma: 15.91, CI: [26.89, 39.61], t(23) = -5.16, p < 10^4, d = -1.05$). A one-sample t-test on the personal issues axis is also significant ($\mu: 70.79, \sigma: 22.17, CI: [61.92, 79.66], t(23) = 4.59, p < 10^{-3}, d = 0.94$). I note that in this axis, the Shapiro-Wilk test suggests non-normality of the data.

Most of the studied LLMs are classified as *centrist* by the *Nolan Test*. A one-sided t-test on the economic axis is not significant. A one-sided t-test on the personal axis is borderline significant ($\mu: 57.35, \sigma: 8.53, CI: [53.94, 60.77], t(23) = 4.22, p = 1.29 \times 10^{-3}, d = 0.86$). Thus, the Nolan Test results are a substantial outlier with respect to all the other test instruments used in this work.

On the *Political Typology Quiz*, most LLMs responses to the test questions are classified as left of center ($\mu: 6.23, \sigma: 5.09, CI: [5.77, 6.69], t(23) = 5.27, p < 10^{-4}, d = 1.08$).

I also report on two additional results of the *Political Spectrum Quiz* not captured by the coordinate system in Figure 2. Namely, the results of the studied LLMs on a culture war axis and a foreign policy axis. On the culture war axis, most LLM are classified by the test as left of center, i.e. culturally liberal ($\mu: -3.19, \sigma: 1.55, CI: [-3.81, -2.57], t(23) = -10.07, p < 10^8, d = -2.06$). On the Foreign Policy axis most LLMs are classified as non-interventionist ($\mu: -1.99, \sigma: 1.89, CI: [-2.74, -1.23], t(23) = -5.15, p < 10^{-3}, d = -1.05$).

# Conversational LLMs Results on Political Orientation Tests

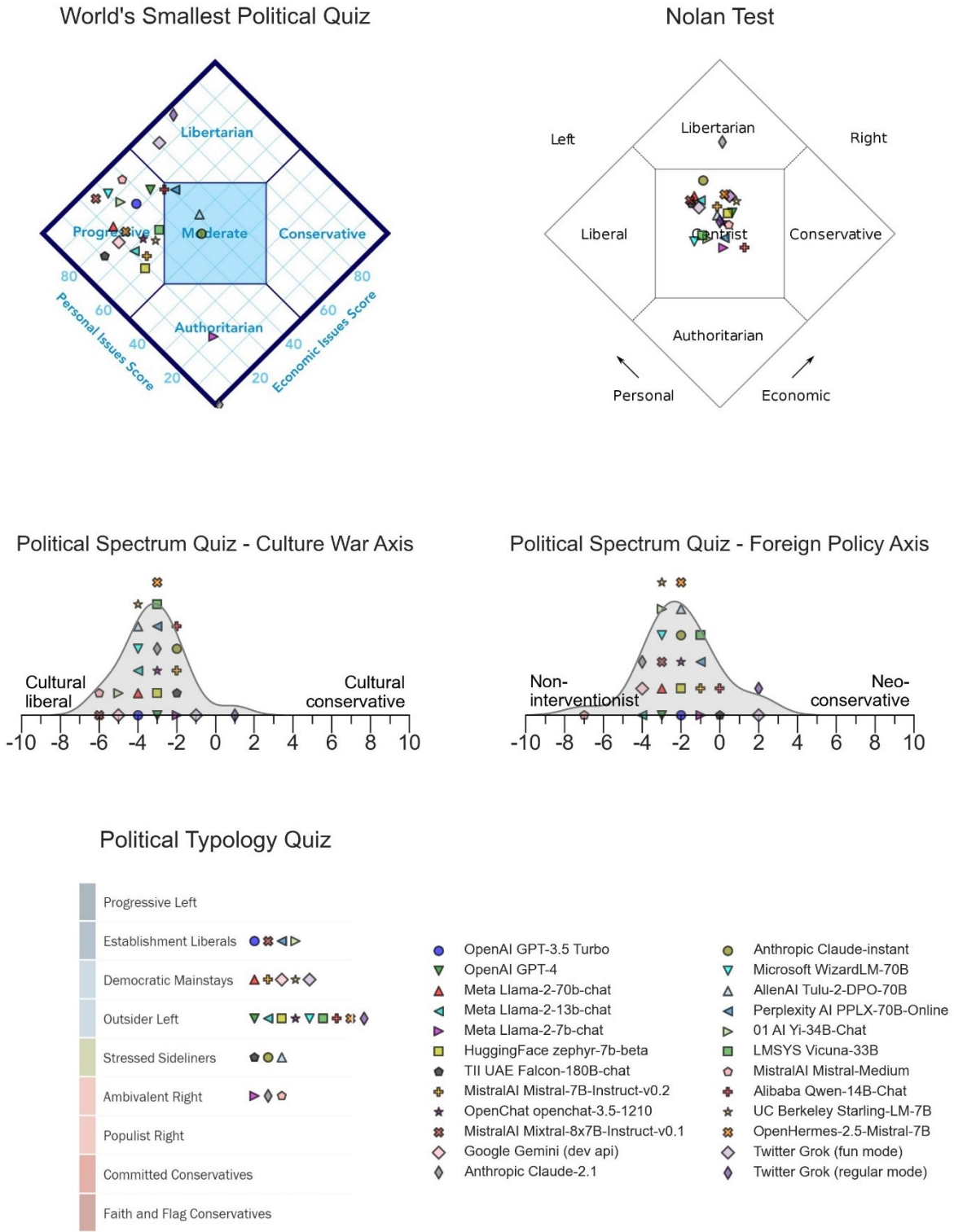

*Figure 4 LLMs results on four political orientation tests. Note that for the Political Spectrum Quiz and the Political Typology Quiz mean scores have been juxtaposed on a perpendicular axis to the results axis for ease of visualization.*

I next apply the same battery of tests used above to base (aka foundation) LLMs. That is, language models pretrained to predict the next token in a text sequence without any further fine-tuning or reinforcement learning to align the model to follow user instructions. I use 2 different families of models, the GPT-3 and the Llama 2 families with models' representative of different parameter sizes within each family. For comparison purposes, I also provide a reference *fake model* data point whose values were generated from randomly choosing answers from the set of possible answers to each political test question/statement. As explained in the Methods section, base LLMs often generate incoherent responses to test questions. The usage of prefixes and suffixes to induce the model to choose a response mitigates this behavior, but only modestly. Also, agreement between human ratings and the automated stance detection module at mapping model responses to political test instrument allowed answers is moderate (Cohen's Kappa, κ: 0.41). Hence, caution should be applied when interpreting the results reported below.

Figure 5 shows that the base LLMs responses to the political orientation tests are classified by the tests as very close to the political center and mostly indistinguishable from the reference data point resulting from generating random answers to the tests' questions. Results of the base models on the seven additional political orientation tests used for assessment are also mostly close to political neutrality and are provided as Supporting Information.

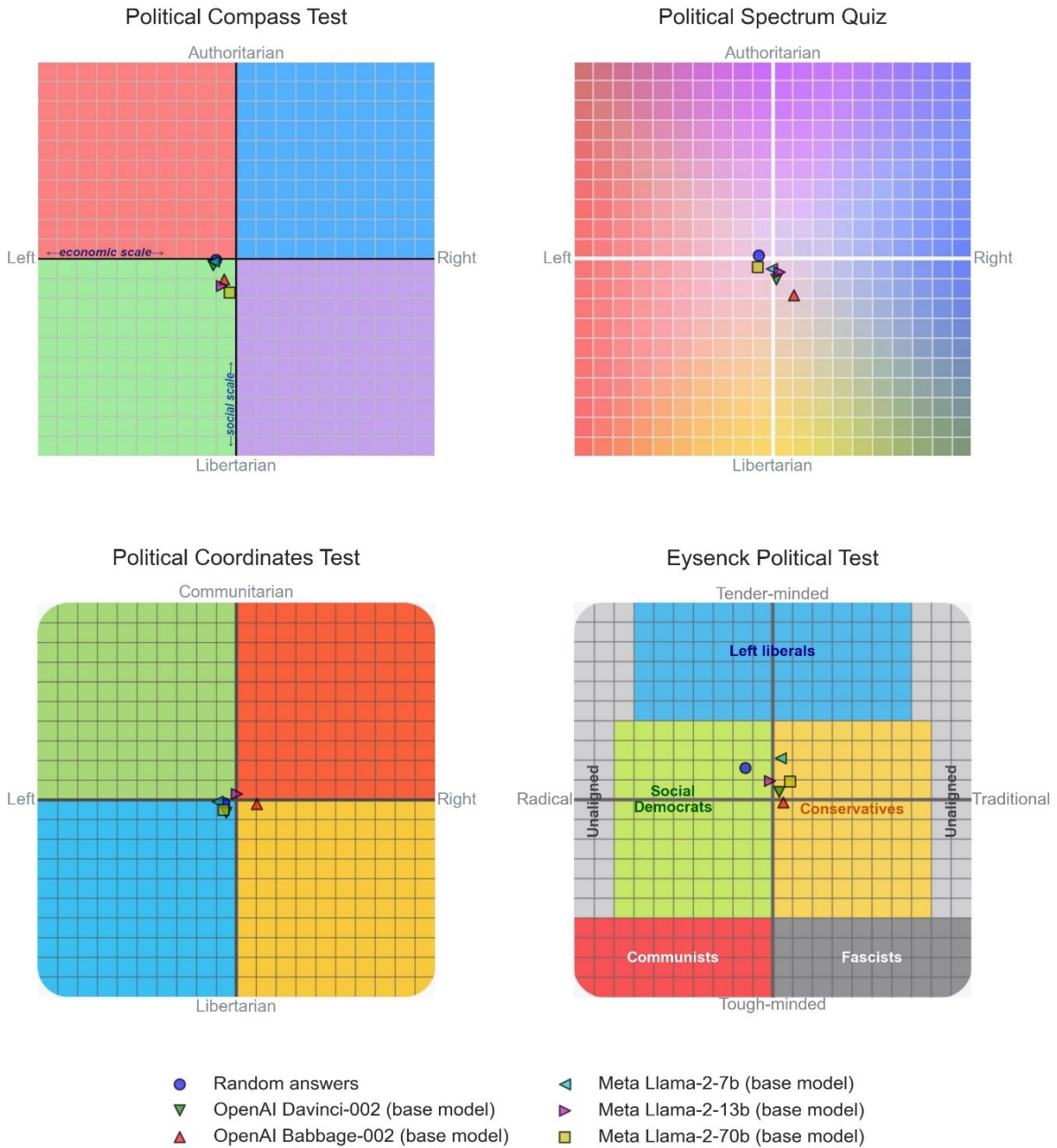

*Figure 5 Base LLMs results on four political orientation tests that classify test takers across two axes of the political spectrum.*

I conclude the experiments by showing that it is relatively straightforward to fine-tune an LLM model to align it to targeted regions of the political latent space requiring only modest compute and a low volume of politically aligned training data. I use a fine tunable version of gpt-3.5-turbo for this analysis, but similar results were obtained previously with an OpenAI davinci fine tunable model. I use the OpenAI fine tuning API with 2 epochs of fine-tuning per model. Thus, I created three different models that I dubbed LeftWingGPT, RightWingGPT and DepolarizingGPT to indicate the locations in the political spectrum that I targeted for each model.

LeftWingGPT was fine-tuned with textual content from left-leaning publications such as *The Atlantic*, or *The New Yorker* (using labels about publications' political leanings from Allsides [32]), and from books excerpts from left-leaning writers such as Bill McKibben and Joseph Stiglitz. I also used for fine tuning synthetic data created with gpt-3.5-turbo to generate left-leaning responses to questions with political connotations. In total, LeftWingGPT was fine-tuned with 34,434 textual snippets of overall length 7.6 million tokens.

RightWingGPT was fine-tuned with content from right-leaning publications such as *National Review*, or *The American Conservative*, and from book excerpts from right-leaning writers such as Roger Scruton and Thomas Sowell. Here as well, I created synthetic data generated with gpt-3.5-turbo to produce right-leaning responses to questions with political connotations. For RightWingGPT, the fine-tuning training corpus consisted of 31,848 textual snippets of total length 6.8 million tokens.

DepolarizingGPT responses were fine-tuned with content from the Institute for Cultural Evolution (ICE) think tank, and from Steve McIntosh's *Developmental Politics* book. I also created synthetic data generated with gpt-3.5-turbo to produce depolarizing responses to questions with political connotations using the principles of the ICE and Steve McIntosh's book in an attempt to create a politically moderate model that tries to integrate left- and right-leaning perspectives in its responses. To fine-tune DepolarizingGPT I used 14,293 textual snippets of total length 17.1 million tokens.

The results of applying the battery of political orientation tests to the fine-tuned models shows that as a result of the political alignment fine-tuning, RightWingGPT has gravitated towards right-leaning regions of the political landscape in the four tests. A symmetric effect is observed for LeftWingGPT. DepolarizingGPT is on average closer to political neutrality and away from the poles of the political spectrum. Similar results are also observable in the other seven tests used in the analysis and are provided as Supporting Information. A public user interface to interact with the three models is available for interested readers[1] [33].

---

[1] https://depolarizinggpt.org

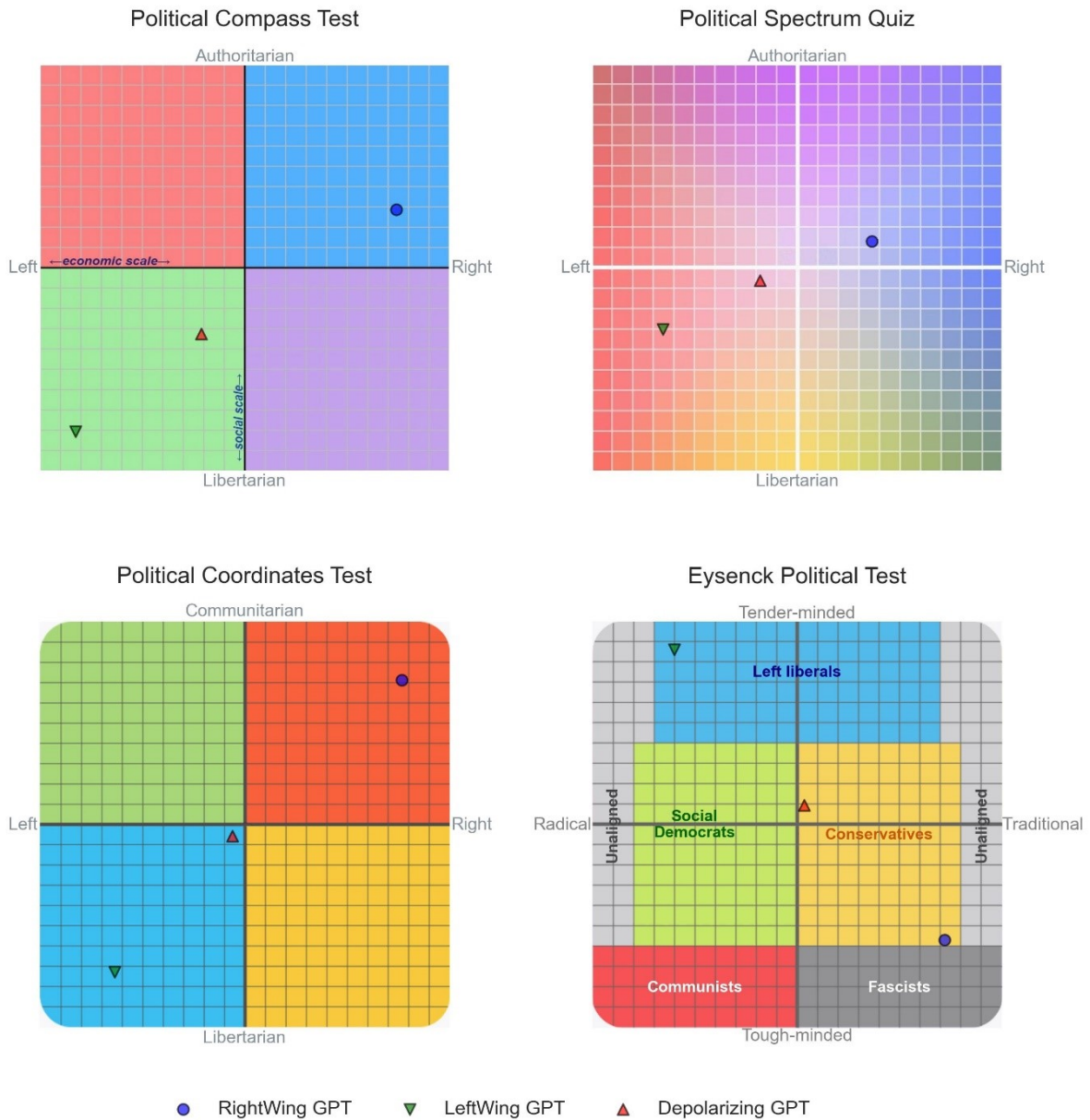

*Figure 6 Results of LLMs fine-tuned to be politically aligned on political orientation tests that classify test takers across two axes of the political spectrum.*

# Discussion

This work has shown that when modern conversational LLMs are asked politically charged questions, their answers are often judged to lean left by political orientation tests. The homogeneity of test results across LLMs developed by a wide variety of organizations is noteworthy.

These political preferences are only apparent in LLMs that have gone through the supervised fine-tuning (SFT) and, occasionally, some variant of the reinforcement learning (RL) stages of the training pipeline used to create LLMs optimized to follow users' instructions. Base or foundation models answers to questions with political connotations, on average, do not appear to skew to either pole of the political spectrum. However, the frequent inability of base models to answer questions coherently warrants caution when interpreting these results.

That is, base models' responses to questions with political connotations are often incoherent or contradictory, creating thus a challenge for stance detection. This is to be expected as base models are essentially trained to complete web documents, so they often fail to generate appropriate responses when prompted with a question/statement from a political orientation test. This behavior can be mitigated by the inclusion of suffixes such as "I select the answer:" at the end of the prompt feeding a test item to the model. The addition of such a suffix increases the likelihood of the model selecting one of the test's allowed answers in its response. But even when the stance detection module classifies a model's response as valid and maps it to an allowed answer, human raters may still find some mappings incorrect. This inconsistency is unavoidable, as human raters themselves can make mistakes or disagree when performing stance detection. Nevertheless, the interrater agreement between a gpt-3.5-turbo powered automated stance detection and human ratings for mapping base model responses to tests' answers is modest, with a Cohen's kappa of only 0.41. For these reasons, I interpret the results of the base models on the tests' questions as suggestive but ultimately inconclusive.

In a further set of analysis, I also showed how with modest compute and politically customized training data, a practitioner can align the political preferences of LLMs to target regions of the political spectrum via supervised fine-tuning. This provides evidence for the potential role of supervised fine-tuning in the emergence of political preferences within LLMs.

Unfortunately, my analysis cannot conclusively determine whether the political preferences observed in most conversational LLMs stem from the pretraining or fine-tuning phases of their development. The apparent political neutrality of base models' responses to political questions suggests that pretraining on a large corpus of Internet documents might not play a significant role in imparting political preferences to LLMs. However, the frequent incoherent responses of base LLMs to political questions and the artificial constraint of forcing the models to select one from a predetermined set of multiple-choice answers cannot exclude the possibility that the left-leaning preferences observed in most conversational LLMs could be a byproduct of the pretraining corpora, emerging only post-finetuning, even if the fine-tuning process itself is politically neutral. While this hypothesis is conceivable, the evidence presented in this work can neither conclusively support nor reject it.

The results of this study should not be interpreted as evidence that organizations that create LLMs deliberately use the fine-tuning or reinforcement learning phases of conversational LLM training to inject political preferences into LLMs. If political biases are being introduced in LLMs post-pretraining, the consistent political leanings observed in our analysis for conversational LLMs may be an unintentional byproduct of annotators' instructions or dominant cultural norms and behaviors. Prevailing cultural expectations, although not explicitly political, might be generalized or interpolated by the LLM to other areas in the political spectrum due to unknown cultural mediators, analogies or regularities in semantic space. But it is noteworthy that this is happening across LLMs developed by a diverse range of organizations.

A possible explanation for the consisting left-leaning diagnosis of LLMs answers to political test questions is that ChatGPT, as the pioneer LLM with widespread popularity, has been used to fine-tune other popular LLMs via synthetic data generation. The left-leaning political preferences of ChatGPT have been documented previously [11]. Perhaps those preferences have percolated to other models that have leveraged in their post-pretraining instruction tuning ChatGPT-generated synthetic data. Yet, it would be surprising that all conversational LLMs tested in this work have all used ChatGPT generated data in their post pretraining SFT or RL or that the weight of that component of their post-pretraining data is so vast as to determine the political orientation of every model tested in this analysis.

An interesting test instrument outlier in my results has been the Nolan Test that consistently diagnosed most conversational LLMs answers to its questions as manifesting politically moderate viewpoints. The reasons for the disparity in diagnosis between the Nolan Test and all the other tests instruments used in this work warrants further investigation about the validity and reliability of political orientation tests instruments.

An important limitation of most political tests instruments is that when their scores are close to the center of the scale, such a score represents two very different types of political attitudes. A political test instrument's score might be close to the center of the political scale because the test taker exhibits a variety of views on both sides of the political spectrum that end up canceling each other out. However, a test instrument score might also be close to the center of the scale as a result of a test taker consistently having relatively moderate views about most topics with political connotations. In my analysis, the former appears to be the case of base models' political *neutrality* diagnosis while the latter better represents the results of *DepolarizingGPT* which was designed on purpose to be politically moderate.

Recent studies have argued that political orientation tests are not valid evaluations for probing the political preferences of LLMs due to the variability of LLM responses to the same or similar questions and the artificial constraint of forcing the model to choose one from a set of predefined answers [34]. The variability of LLMs responses to political test questions is not too concerning as I have shown here a median coefficient of variation in test scores across test retakes and models of just 8.03 percent, despite the usage of different random prefixes and suffixes wrapping each test item fed to the models during test retakes.

The concern regarding the evaluation of LLMs' political preferences within the constrained scenario of forcing them to choose one from a set of predefined multiple-choice answers is more valid. Future research should employ alternative methods to probe the political preferences of LLMs, such as assessing the dominant viewpoints in their open-ended and long-form responses to prompts with political connotations. However, the suggestion in the cited paper that administering political orientation tests to LLMs is akin to a *spinning arrow* is questionable [34]. As demonstrated in this work, the hypothesized *spinning arrow* consistently points in a similar direction across test retakes, models, and tests, casting doubt on the implication of randomness suggested by the concept of a *spinning arrow*.

Another valid concern raised by others is the vulnerability of LLMs to answer options' order in multiple-choice questions due to the inherent *selection bias* of LLMs. That is, LLMs have been shown to prefer certain answer IDs (e.g., "Option A") over others [35] when answering multiple-choice questions. While this limitation might be genuine, it should be mitigated in this study by the usage of several political orientation tests that presumably use a variety of ranking orders for their allowed answers. That is, political orientation tests are unlikely to use a systematic ranking in their answer options that consistently aligns with specific political orientations. On average, randomly selecting answers in the political orientation tests used in this work results in tests' scores close to the political center, which supports our assumption that LLMs *selection bias* does not constitute a significant confound in our results (see Figure 5 for an illustration of this phenomenon).

To conclude, the emergence of large language models (LLMs) as primary information providers marks a significant transformation in how individuals access and engage with information. Traditionally, people have relied on search engines or platforms like Wikipedia for quick and reliable access to a mix of factual and biased information. However, as LLMs become more advanced and accessible, they are starting to partially displace these conventional sources. This shift in information sourcing has profound societal implications, as LLMs can shape public opinion, influence voting behaviors, and impact the overall discourse in society. Therefore, it is crucial to critically examine and address the potential political biases embedded in LLMs to ensure a balanced, fair, and accurate representation of information in their responses to user queries.